\documentclass[11pt]{article}
\pdfoutput=1
\usepackage{hyperref}
\begin{document}
\title{The Wet-Dog Shake}
\author{Andrew Dickerson, Grant Mills, Jay Bauman, Young-Hui Chang, David Hu\\
\\\vspace{6pt} George W. Woodruff School of Mechanical Engineering, \\ Georgia Institute of Technology, Atlanta, GA 30332, USA}
\maketitle
%% The abstract (in this file, and that submitted as text to arXiv) should include the exact phrase
%% "fluid dynamics video" or "fluid dynamics videos"
\begin{abstract}
The drying of wet fur is a critical to mammalian heat regulation. In this fluid dynamics video, we show a sequence of films demonstrating how hirsute animals to rapidly oscillate their bodies to shed water droplets, nature's analogy to the spin cycle of a washing machine. High-speed videography and fur-particle tracking is employed to determine the angular position of the animal's shoulder skin as a function of time.  X-ray cinematography is used to track the motion of the skeleton.  We determine conditions for drop ejection by considering the balance of surface tension and centripetal forces on drops adhering to the animal. Particular attention is paid to rationalizing the relationship between animal size and oscillation frequency required to self-dry.\end{abstract}
% main text
%\section{Introduction}

%
\end{document}